\documentclass[preprint,showpacs,preprintnumbers,amsmath,amssymb,floats]{revtex4-1}
\usepackage{blindtext}
\usepackage{graphicx}
\usepackage[english]{babel}
\usepackage{amsmath}
\usepackage{amssymb}
\usepackage{color}

\newcommand{\be}{\begin{eqnarray}}
\newcommand{\ee}{\end{eqnarray}}
\usepackage[toc,page]{appendix}

\begin{document}
\title{Quantum-criticality in twisted  bi-layer graphene}

\author{C. M. Varma}
\affiliation{
Department of Physics, University of California, Berkeley, CA. 94720 $^+$\\
Department of Physics, University of California, Riverside, CA. 92521 
} 
\thanks{ Emeritus, $^+$ Visiting Scholar}
\date{\today}
\begin{abstract} 

Transport experiments in twisted bilayer graphene (TBG) show a fan-like region near integer fillings with a resistivity linear in temperature  down to the lowest temperature measured. This suggests quantum-critical points at the boundary to long-range ordered phases. The particular order proposed by Blutinck et al. for twisted bi-layer graphene (TBG) is a loop-current order at the carbon length-scale together with modulations on the moir\'e length scale. This is shown to be the ground state of a  xy model with translational symmetry and time-reversal broken. 
 Here, this is extended to derive a model for quantum-critical properties.  The kinetic energy operator for the model and the coupling of fermions to the fluctuations of the xy model are derived. The previously derived universal properties in the quantum-critical region of such a model, leading to a marginal Fermi-liquid, irrespective of the underlying microscopics is briefly reviewed. The properties include the resistivity and various other transport properties with and without applying a magnetic field and the instability of the quantum fluctuating state to superconductivity in d-wave symmetry.
    \end{abstract}

\maketitle

\section{Introduction}
The fabrication of twisted multilayer Graphene (TBG) \cite{Cao:2018il, Cao:2018ef, EAndrei:2021}
allows electron densities in the bands close to the chemical potential to be continuously varied as well as the ratio of electronic interaction energies   to the kinetic energy. This has led to the discovery of a variety of phases. Of special interest in this paper are regions of the phase diagram in which the insulator to metal transition is accompanied by a characteristic fan-like quantum-critical region in which the resistivity is linear in temperature, followed at lower temperature with a superconducting region \cite{Cao:2018ef, Efetov2019}. On suppressing superconductivity with a magnetic field $H$, a region of resistivity linear in $H$ is also found \cite{Cao:2020, Efetov2019}. Such properties are found in all compositions near integer filling in TBG accompanied by a superconducting dome in each case, but not at half-filling. These properties are the same as the phenomena in cuprates near their loop-current order \cite{Kaminski-diARPES, simon-cmv, Bourges-rev}, in heavy-fermions near their antiferromagnetic quantum-critical point (AFM-QCP) \cite{LohneysenRMP2017}, and in Fe-based anti-ferromagnets \cite{Shibauchi_Rev:2014}, also near their AFM-QCP. 

Theories for Quantum-criticality have been investigated extensively in the last few decades \cite{LohneysenRMP2017, VarmaRMP2020}. At a phenomenological level, agreement with a wide variety of experiments has always required that the critical fluctuations obey the Marginal Fermi-liquid paradigm \cite{CMV-MFL}, which is that the critical fluctuations have their absorptive part at frequency $\omega$ proportional to $\tanh(\omega/2T)$ or equivalently that they decay as $1/\tau$, where $\tau$ is the imaginary time periodic in $2\pi/T$. The momentum-dependence of the fermions  due to scattering of the critical fluctuations should be 
  un-important, either due to the locality of the fluctuations themselves or due to the combined effects in the product of their momentum dependence and the coupling function to fermions \cite{ASV2010}. (For comparison, Fermi-liquids have fluctuations which decay as $1/\tau^2$ and the spatial-dependence are similar to the time-dependence. The fermion self-energy is a similar function of energy and deviation of momentum from the Fermi-surface in a typical Fermi-liquid.)  
  
The statistical mechanical model which is derivable from the microscopic  model for the loop-current order in cuprates and for the AFM in-plane order or incommensurate Ising order in the heavy fermions or Fe-based compounds   is the quantum-xy model \cite{Aji-V-qcf1, Varma_IOPrev2016} coupled to fermions.  The essential part of the latter is  that the
fluctuations couple to the fermion charge currents \cite{ASV2010} or spin currents \cite{Varma_IOPrev2016}.  This form of the coupling is also
important for the symmetry of favored superconducting state. Such a model has been solved by quantum-monte-carlo \cite{ZhuChenCMV2015, ZhuHouV} and by renormalization group \cite{Hou-CMV-RG} and has precisely the properties of the Marginal Fermi-liquid paradigm. It is the purpose of this paper to show how  the model proposed by Blutinck et al for Moir\'e  TBG \cite{Zaletel2020} maps to the quantum xy model coupled to fermions. 
\section{Twisted bi-layer Graphene (TBG)}

\begin{figure}[h]
 \begin{center}
 \includegraphics[width= 0.6\columnwidth]{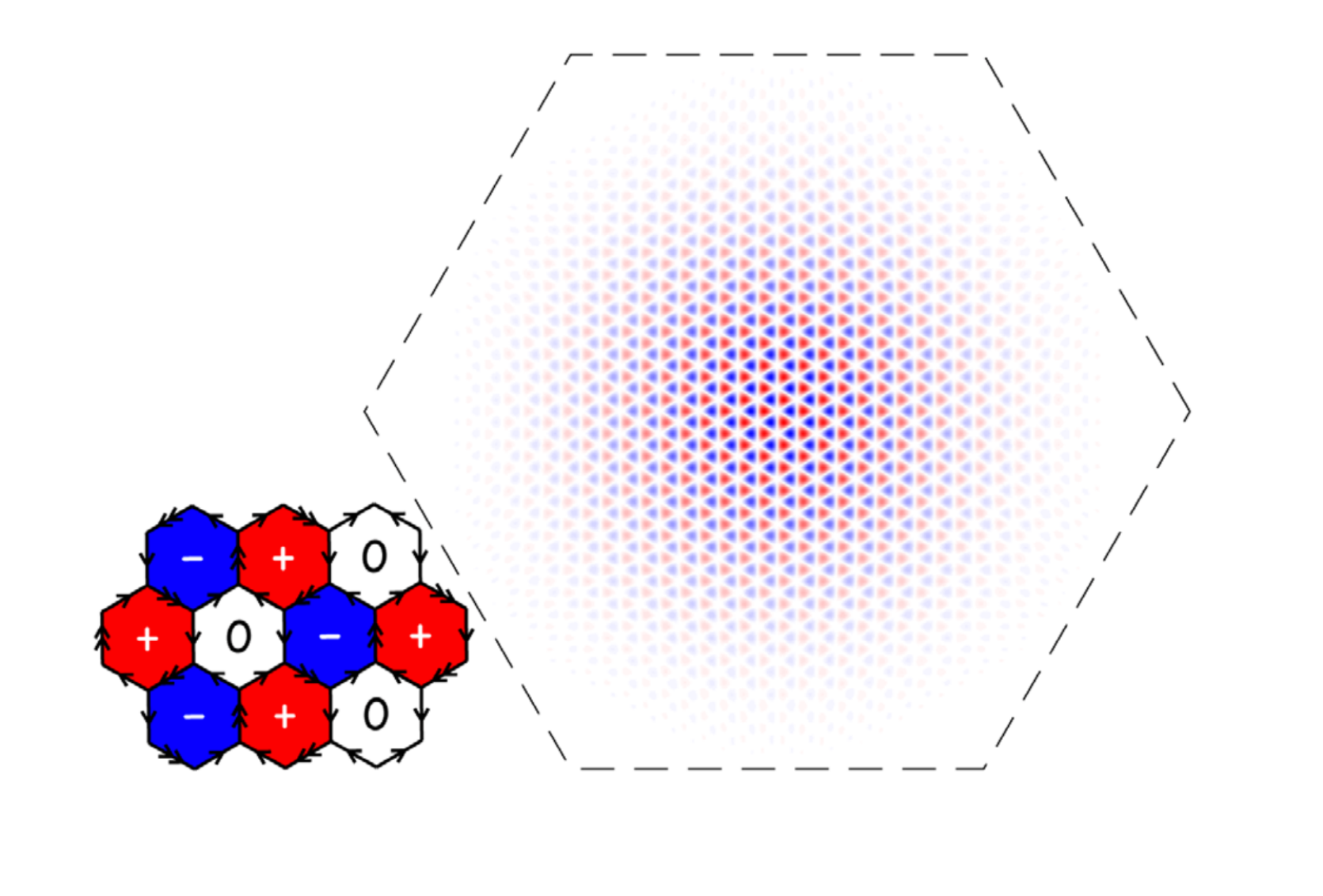}
 \end{center}
\caption{Order parameter proposed in Ref. \cite{Zaletel2020} in twisted bi-layer graphene. The main figure is on the Mo\`ire length scale but it is decorated by order on the hexagonal graphene length scale with modulation of its amplitude on the larger length scale. The decoration, shown in the smaller figure, is an enlarged $\sqrt{3} \times \sqrt{3}$ unit-cell which results in six hexagons with flux alternating in direction around a hexagon which necessarily has no flux. Such an order was first proposed  in graphene and bi-layer graphene  \cite{Zhu-A-V2013} in a model with on-site and neighboring site repulsive interactions in a mean-field calculation similar to that in \cite{Zaletel2020}.}
 \label{Fig:TBG_order}
\end{figure}

A time-reversal odd state has also been predicted by mean-field calculations \cite{Zaletel2020} on a reduced model for TBG \cite{Bis_Mac} and verified in sign free Monte-Carlo calculations \cite{Berg2021} at particle-hole symmetry on the same model. The order proposed is at the graphene lattice scale 
with a modulation on the moir\'e scale.  On the graphene scale, the order consists of a flux pattern in which there is a $\sqrt{3} \times \sqrt{3}$ increase of the unit-cell of graphene with a hexagon with no flux surrounded alternately by three hexagons with positive and with negative flux spontaneously generated by loop-currents at the perimeters of the hexagons -See Fig. (\ref{Fig:TBG_order}) reproduced from Fig 1 of Ref. \cite{Zaletel2020}. There is necessarily a modulation of the amplitude of the order near the boundary of the enlarged TBG lattice scale. On the graphene scale, exactly the same flux pattern was proposed as a stable state about a decade earlier for bi-layer graphene with short-range interactions -See Fig. (7) of \cite{Zhu-A-V2013}. As discussed there, this is the simplest flux pattern possible on a hexagonal lattice because of the frustration for an AFM- Ising model for spins (or fluxes) on a triangular lattices.

 The model proposed for TBG \cite{Zaletel2020} is suitable for a theory of the classical transition to the state of lower symmetry and fluctuations in the classical regime near such a transition. In this paper, we add to this model the required terms to give quantum-fluctuations. Such  quantum terms provide the kinetic energy and are formed from the generator of rotations among equivalent states of order. They must also be represented by a loop-current pattern. The procedure to generate such  terms is modeled on the theory for the cuprates \cite{ASV2010}.  It is worthwhile noting that incommensurate charge density wave order, if it exists, also belongs to the class whose criticality is of the xy class. 

\subsection{Order Parameter}

We will primarily consider the order parameter and the fluctuations on the spatial scale of the graphene lattice. The order and fluctuations on the  much larger length scale can be treated similarly but the thermodynamic strength and the amplitude of the fluctuations for the latter are smaller than the latter by the inverse ratio of the hexagonal  moi/'re lattice to the hexagonal graphene lattice. This assumes that the order at the moi\'re scale and the graphene scales are coincident in temperature for any given doping.

 The low energy states in the graphene lattice lie near the K, K' points of the Brillouin zone; there are three pairs of such points lying at $\pm \pi/3$ with respect to each other. As is customary, we define the Pauli matrix  $\sigma$ is in the sub-lattice space and $\tau$ in the valley space and $\omega$ the spin-space.  We drop the spin-index; it plays no role.  The order parameter proposed by Bultinck et al. \cite{Zaletel2020} (and Zhu et al. \cite{Zhu-A-V2013}) is

\begin{eqnarray}
\label{OP}
{\bf \Psi} \equiv < \psi^+ \sigma_y (\tau_x \hat{x} + \tau_y \hat{y}) \psi >.
\end{eqnarray}
 This order parameter breaks the valley conservation which is a  $U(1)$ symmetry in the continuum but a six fold symmetry in the lattice. It also breaks time-reversal symmetry ${\mathcal{T}} = \tau_x C$ under which ${\bf \Psi}_{(x,y)} \to - {\bf \Psi}_{(x,y)}$, i.e. $K,K' \to -K,-K'$. It preserves the three fold symmetry in rotation of $\pi/3$:, i.e in the operator $e^{i 2\pi/3 \sigma_z \tau_z}$ but breaks inversion $\tau_x \sigma_x {\bf \Psi}_{(x,y)} \to {\bf \Psi}_{(x)}, - {\bf \Psi}_{(y)}$.
 Note that there is no symmetry associated with the sub-lattice $\sigma_z: A \to B$.
 
 As discussed below, we can represent the order parameter in the continuum simply as
 \begin{eqnarray}
 {\bf \Psi} = |{\bf \Psi}| e^{i \phi}
 \end{eqnarray}
 where $|{\bf \Psi}|$ is its amplitude and $\phi$ is its phase. The simplest Hamiltonian for the fluctuations of $\phi$ is the quantum xy-model which must however be supplemented by coupling to fermions to be relevant to our problems. This is discussed in the next section.
 
 It is interesting to derive the loop-current patterns in real space from the order parameter. To do so, we first write the Bloch wave-functions in the ordered states at the points K and K' in the Brillouin zone:
 \begin{eqnarray}
 \label{psi}
 \psi_{\sigma_z, \tau_z}({\bf r}) = \sum_{i =1,2,3} e^{\tau_z {\bf K}_i \cdot({\bf r} + \frac{\sigma_z}{2} \hat{y})}
 \end{eqnarray}
 More explicitly the two coupled wave-functions at $K, K'$, in the ordered state are
 \begin{eqnarray}
 \label{Kwavefn}
 \psi_1 &=& \big(\psi_{A,K}({\bf r}) + i r e^{i \phi} \psi_{B,K'}({\bf r})\big)/\sqrt{1+r^2},\\
 \psi_2 &=& \big(\psi_{B,K}({\bf r}) - i r e^{i \phi} \psi_{A,K'}({\bf r})\big)/\sqrt{1+r^2}.
 \end{eqnarray}
$r$ accounts for the magnitude of the order parameter and $\phi$ is the rotation angle of the ${\bf \Psi}$ with respect to the x-axis. Note that the rotation of $\phi$ by $n \pi/3$ is equivalent to rotation in $K,K'$ space. $a$ is the carbon-carbon distance and ${\bf K, K}'$ are the reciprocal vectors of the new Brillouin zone reduced by $\sqrt{3}$ from those of the graphene lattice: ${\bf K}_1 = \frac{4 \pi}{3\sqrt{3} a} \hat{x}, {\bf K}_2 = \frac{4 \pi}{3\sqrt{3} a} \big(- \frac{\hat{x}}{2} + \frac{\sqrt{3}\hat{y}}{2}\big),
 {\bf K}_3 = \frac{4 \pi}{3\sqrt{3} a} \big( -\frac{\hat{x}}{2} - \frac{\sqrt{3}\hat{y}}{2}\big)$. The wave-functions at any point of the zone may be constructed from (\ref{Kwavefn}).
 
 Given the wave-functions, the current in the bonds and the equivalent magnetization enclosed by the bonds can be calculated. A pattern with such a broken symmetry is found in Fig. (7) in Ref. \cite{Zhu-A-V2013} and in Fig. (1) in Blutinck et al. \cite{Zaletel2020} reproduced above for TBG on the carbon length scale in any Moire-period. This represents one of the six patterns possible; the others are obtained by a $\pi/3$-rotation. In the quantum (and classical) fluctuation region, there are spatial and temporal fluctuations of the six patterns. If the six fold anisotropy, which is irrelevant both classically and quantum-mechanically is ignored, the fluctuations are that of an xy model. 

 It is worthwhile writing down the real space Hamiltonian whose eigenstates are equivalent to those discussed above.
 It is simply
 \begin{eqnarray}
 \label{H0}
     H_0 = (t+  r e^{i\phi})\sum_{i, \delta_{\mu}}a^+_i b_{i + \delta_{\mu}} + H.C.
 \end{eqnarray}
 $a_i, b_i$ are annihilation operators on the $A$ and the $B$ sub-lattice in unit-cell $i$. $t$ is the real one-particle transfer matrix element and $r$ is the amplitude of the loop-current order and $\phi$ is the phase. They have identical meaning as the parameters in Eq. (\ref{psi}).
 The current pattern for the occupied eigenstates of (\ref{H0}) give the pattern with an unit-cell enlarged by $\sqrt{3} \times\sqrt{3}$. It has the periodic pattern of a hexagon with zero-flux, surrounded by six hexagons with alternate flux. This is the best that can be done with currents on nearest neighbor complex transfer integrals, because the flux is an Ising object and the centers of the hexagons lie on a triangular lattice so that to prevent frustration, a hexagon in the enlarged lattice must always have zero flux. 
    
 \subsection{Generator of Rotations and its coupling to fermions}
 
 \begin{figure}[h]
 \begin{center}
 \includegraphics[width= 0.5\columnwidth]{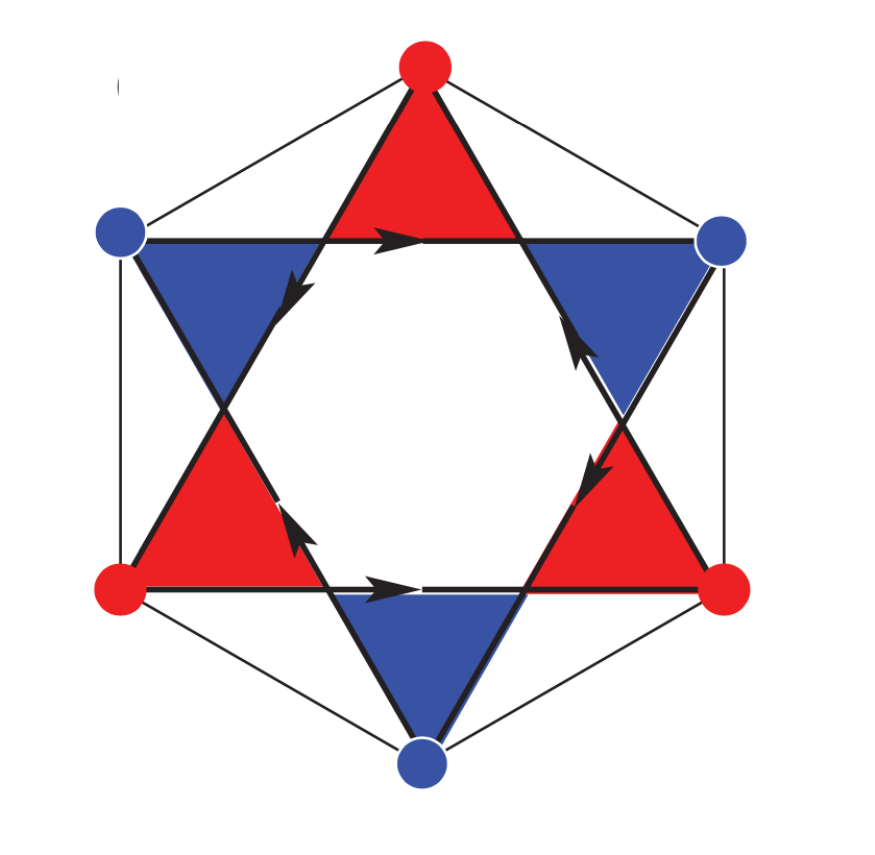}
 \end{center}
\caption{The generator of six-fold rotations of the principal order parameter (on the graphene length scale) shown in Fig. (\ref{Fig:TBG_order}). This model must be distinguished from the Haldane quantized Hall effect current pattern \cite{Haldane1988} in a hexagonal lattice
in which current loops are in the same direction on the two triangular sub-lattices while here they are in opposite directions and break inversion symmetry.}
 \label{Fig:Rotator}
\end{figure}

 To generate local spatial (and temporal) fluctuations of the flux patterns, one must find  the angular momentum operator $L_z$ which rotates among the 6 different  possible flux patterns.  A look at Fig. (\ref{Fig:TBG_order}) reveals that this is also the translation operator which in any of the patterns switches the zero flux hexagons with the up or down flux hexagons. This is accomplished by reversing three of the currents on the alternate boundaries of the hexagon with zero flux. The real space current pattern corresponding to $L_z$ was derived by Zhu et al. \cite{Zhu-A-V2013} in Fig. (2b) of that paper and is reproduced here as Fig. (\ref{Fig:Rotator}). 
It differs from the topological  Haldane pattern \cite{Haldane1988} which has currents flow in the same direction in the triangles connecting the same sub-lattice sites in each of the  hexagons. Note that in our case, this pattern would also form a $\sqrt{3} \times \sqrt{3}$ lattice if ordered.

In the ferromagnetic  xy model, the amplitude of the fluctuations is established well above the critical point and is nearly constant while the angular variations have the critical fluctuations. For the antiferromagnetic model, the important  fluctuations are of the orientations of the order parameter around the six equivalent possible AFM Bragg vectors. 6-fold anisotropy is irrelevant  both the classical and the quantum xy model. So in the fluctuation region, instead of the 6 equivalent Bragg vectors of fixed length, we should consider an order parameter  about a circle of fixed length with the fluctuations of the angle of the order parameter at any point on the circle. The generator of rotations, which must be determined for the quantum kinetic energy, may be constructed from that which gives the infinitesimal rotation of the continuum xy model. Below we consider such a rotation and its coupling to corresponding fermions. In real space, this is equivalent to fluctuations keeping the radial vector of the order parameter fluctuation about  $\sqrt{3} a$ and rotating its direction by $\theta$ so that the six-fold orientations of the red and the blue hexagons around any white hexagon in  Fig. (\ref{Fig:TBG_order}) changes to a a continuous color change which is an admixture of red and blue.  In momentum space, this is equivalent to considering the order parameter fluctuations to be around a magnitude with ${\bf Q}_0$ in a circle, and considering tangential infinitesimal rotations at a momentum $\delta{\bf q} \bot {\bf Q}_0$ with magnitude which for infinitesimal rotations is $\delta(\theta({\bf q})) = (\delta({\bf q})/{\bf Q})$ . Correspondingly the coupling to fermions is also at momentum ${\bf Q}_0 + \delta{\bf q}$. We will concentrate only on the angular fluctuations and not concern ourselves with the solid to the so called hexatic criticality \cite{HalpNelson1979}.

 \subsection{Coupling to fermions of the phase field}

 We derive the coupling of fermions to the collective loop-currents which have the symmetries described above; i.e.  the coupling of the fermions to  the fluctuations $\delta \phi({\bf r})$. As is natural the coupling is to fermion-currents. In order to minimize subscripts, we will use notation which is more suitable for the ferromagnetic xy model and indicate at the end how the results translate to the AFM model.
  
 We may write the variations $\delta \phi({\bf r})$ generally in terms of their orthogonal components $\delta \phi_x, \delta \phi_y$. The eigenvalues of the Hamiltonian are invariant to $(\phi_x, \phi_y)$ as long as they lie on a circle. This is equivalent to neglecting anisotropies, i.e. to having realization for criticality of the xy-model. We may then consider radial variations of $\delta {\vec{\phi}}_r({\bf r})/\delta {\bf r} = \nabla \phi({\bf r})$ and tangential variations 
 $\delta {\vec{\phi}}_t({\bf r}))/\delta {\bf r} = \nabla \times \phi({\bf r})$. The latter may be explicitly written as 
\begin{eqnarray}
\left(\begin{array}{ccc}\delta \phi_{t,x}({\bf r}) \\\delta \phi_{t,y}({\bf r})\end{array}\right) = \frac{1}{|\phi|}\left(\begin{array}{ccc} - \phi_y \\ \phi_x\end{array}\right) \delta \theta({\bf r}).
\end{eqnarray}
$\delta \theta({\bf r})$ is the rotation of the phase $\phi({\bf r})$ about the $z$-axis, i.e. due to the generator of rotations of the field $\phi({\bf r})$.  $|\phi| = \sqrt{\phi_x^2 + \phi_y^2}.$ For the longitudinal variations $\nabla \phi({\bf r}) = \nabla \theta({\bf r}).$

 Let us consider the coupling to the partially filled band whose creation and annihilation operators are denoted by $c_{\bf k}^+, c_{\bf k}$ and has eigenvalues are given by diagonalizing Eq. (\ref{H0}), and which  for $r \phi << t$ are $\epsilon(k_x a + \phi_x, k_y a + \phi_y)$. This form is important as it shows we have a gauge theory coupling, which ensures that the currents produced by a variation $\delta \theta(r)$ are conserved at all lattice point. 
 
 Let us consider slow variations of $\delta \phi_x$ and $\delta \phi_y$ so that we can expand,
 \be
 \epsilon(k_x a + \phi_x, k_y a + \phi_y) = \epsilon(k_x a, k_y a ) + \frac{\partial \epsilon}{\partial k_x a} \delta \phi_x ({\bf r}) +  \frac{\partial \epsilon}{\partial k_y a} \delta \phi_y ({\bf r}) + ....
 \ee
 Here  we have noted that taking the derivative of $\epsilon$ with respect to $\phi_x, \phi_y$ is equivalent to taking the derivative with respect to $k_x a, k_y a$, i.e. proportional to the velocity. Also, in the Fourier transforms the leading term in $\phi_x \delta \theta(q)$ is proportional to $q_x = (k-k')_x$, etc. 
 
 Consider as above  variations $\delta \phi ({\bf r})$ both along and perpendicular to ${\bf r}$. They lead respectively to  two terms in the Hamiltonian $H' = H_r' +H_t'$ giving the coupling of fermions to the collective fluctuations $\delta \theta({\bf q})$: 
\begin{eqnarray}
 H'_r &= &\sum_{{\bf k,q}}  \theta_r  \gamma_r({\bf k, k+q})~ c_{{\bf k+q}}^+ c_{\bf k} + H.C. \\
\gamma_{r} &= &(\epsilon_{{\bf k+q}} - \epsilon_{{\bf k}}) \approx {\bf q}\cdot {\bf v}({\bf k}_F)
\end{eqnarray}
and
\begin{eqnarray}
 H'_t &= &\sum_{{\bf k,q}}  \theta_t ~{\bf q} \cdot \gamma_t({\bf k, k+q}) c_{{\bf k+q}}^+ c_{\bf k} + H.C. \\
\gamma_t({\bf k,k')} &=&  \big( - {\bf v}_y( {{\bf k'}}) \hat{\bf x} + {\bf v}_x({{\bf k}})\hat{\bf y} \big).
\end{eqnarray}
We see that the symmetry of the coupling to angular fluctuations is given by
\begin{eqnarray}
\gamma_t(k, k') \propto  ({\bf k} \times {\bf k'}) 
\end{eqnarray}
This coupling is consistent with the fact that a generator of rotations of $\theta$ should couple only to the angular momentum of fermions $c^+_{\bf k} ({\bf r} \times {\bf p}) c_{\bf k'}$. 

For the antiferromagnetic model, the only difference is that the momentum transfer ${\bf q} = {\bf k'-k}$ is measured from the Bragg-vector of the
putative transition ${\bf Q}_i$. For the critical fluctuations for which the six-fold anisotropy is irrelevant, they are to be considered as lying  in a circle. As in all critical phenomena, this irrelevance amounts in real space to considering a region much larger than the lattice constant.

These results are important for two reasons. They give the proper coupling to fermions currents which are integrated over to generate a term in the effective Hamiltonian or the action for  the fluctuations. It turns out that the term generated through coupling to $L_z = i \frac{\partial}{\partial \theta}{\hat{\bf z}}$  by integrating over the  $<L_z ({\bf r}, \tau)L_z(0,0) >$ correlations is irrelevant to the action for the fluctuations \cite{ZhuHouV} . Such correlations, as further explained in the next section, are actually the critical correlations.  The $<\nabla \theta ({\bf r}, \tau) \nabla \theta (0,0)>$ correlations are not critical but they provide coupling to fermion currents ${\bf J}$. Integrating over the $<{\bf J}(({\bf r}, \tau) {\bf J}(0,0)>$ correlations do provide a relevant term in the action for the collective fluctuations. These matters are further explained in the next section. 

 \section{\bf Model for quantum-criticality}
 
 As has been derived above, neglecting the (irrelevant) six fold anisotropy in both TBG and TDC, the order parameter fluctuations are described by
 \begin{eqnarray}
 {\bf \Psi} = |\Psi| e^{i \phi({\bf r})}.
 \end{eqnarray}
  The quantum Hamiltonian necessary to describe the fluctuations includes the kinetic energy in terms of the effective angular momentum variable which has the right commutation relation with ${\bf \Psi}$ so that it provides the generator of rotations for $\phi(r)$ in the plane. In the Hamiltonian formulation, this is the operator $L_{\bf {\hat{z}}} \equiv i |\Psi| \frac{\partial}{\partial \theta} {\hat{\bf z}}$, to which the fermions couple with the coupling functions $\gamma_t$ derived above. It also must include the term generated from the coupling of the fluctuations to the fermions after eliminating the relevant correlation function of the fermions. Since as shown above, the coupling to fermions is to their currents or angular momentum, appropriately defined in each microscopic case, the elimination involves the current or angular momentum currents. These generate frequency or time-dependent contributions. So, it is best to represent the fluctuation Hamiltonian in terms of a Lagrangian. 
 The procedure has been described before \cite{Aji-V-qcf1,Aji-V-qcf3}. Summarizing it, the action is
  \be
  \label{modelxy}
     S = S_{qxy} + S_{c-f}.
     \ee 
     where the first term contains the potential and kinetic energy of the fluctuating variable $\theta({{\bf x}, \tau})$, where ${\bf x}$ is the two-dimensional space co-ordinate and $\tau$ is the imaginary time periodic in $2\pi/T$.
     \be
    S_{qxy} &=&-K_0 \sum_{\langle {\bf x, x}' \rangle} \int_0^{\beta} d \tau \cos(\theta_{{\bf x}, \tau} - \theta_{{\bf x}', \tau}) \nonumber \\
& +& \frac 1 {2E_c} \sum_{{\bf x}} \int_0^\beta d \tau \left( \frac{d \theta_{{\bf x}}}{d\tau}\right)^2.
\ee
$K_0$ is the Josephson coupling of the phase variables and $E_c$ is the moment of inertia for the angular momentum of the fluctuations, which are the generator of the rotations of $\theta({\bf x}, \tau)$ The latter is in terms of the operator $L_{\hat{\bf z}}$ we introduced in the Hamiltonian representation for both physical cases derived above. 

$S_{c-f}$ is the contribution to the action of the fluctuations obtained by integrating over the fermion current or angular momentum correlations. 
In Monte-carlo calculations \cite{ZhuHouV}, it has been shown that the operator introduced though the angular momentum correlations is irrelevant in determining the critical correlation function.  That through the fermion-current correlation, i.e. through the coupling function $\gamma_r$ is relevant. The resulting term in the action $S_{c-f}$ is both easily written and physically transparent in momentum and frequency space. It has the Caldeira-Leggett  \cite{CaldeiraLeggett} form, though the physical basis is different:
\begin{eqnarray}
 \label{H'2}
 S_{c-f} = \sum_{{bf q, \omega}} \alpha |\omega| q^2 |\theta(q,\omega)|^2.
  \end{eqnarray}
  This form comes about because $\nabla \theta$ which after Fourier transform is ${\bf q} \theta$ couples to the fermion current ${\bf J}$. This generates a coupling  $\propto  q^2 ~<JJ>(q, \omega)$. The imaginary part of the current-current correlation is just $|\omega| \alpha$, where $\alpha$ is the dimensionless conductivity in the limit $\omega \to 0, T \to 0$ in the asymptotic quantum-critical region. The Fourier transform of this in real space and imaginary time space looks much more formidable, although it is necessary for Monte-carlo calculations.
\be
\label{c-f}
S_{c-f} &=& \frac{\alpha}{4\pi^2} \sum_{\langle{\bf x, x}'\rangle} \int d \tau  d\tau' \frac {\pi^2}{\beta^2} \frac {\left[(\theta_{{\bf x}, \tau} - \theta_{{\bf x}', \tau})  -(\theta_{{\bf x}, \tau'} - \theta_{{\bf x}', \tau'}) \right]^2}{
\sin^2\left(\frac {\pi |\tau-\tau'|}{\beta}\right)}
 \ee
 
 The correlation function of the action of (\ref{modelxy}) have been obtained in extensive quantum-Monte-carlo calculations \cite{ZhuChenCMV2015, ZhuHouV} and re-obtained analytically  \cite{Hou-CMV-RG} by renormalization group at a level similar to the solution of the classical xy model by Kosterlitz \cite{Kosterlitz1974}. The results are summarized in an Appendix in Ref. \cite{Maebashi-Kubo} and the application to transport and thermodynamic properties given in Ref. \cite{VarmaRMP2020}. The essential of the critical fluctuations is that they ar e of a product form as a function of frequency $\omega$ and momentum ${\bf q}$. The omega dependence has $\omega/T$ scaling with a cut-off at high frequencies given by the microscopic model. 
 
 In the quantum-critical problems of the materials considered in this paper, the primary quantities that are measured are the temperature dependence of the resistivity and the field dependence of the resistivity. The calculation of such properties is extensively discussed through the  solution of the vertex in the Kubo equation for electrical, thermal and thermo-electric  transport \cite{Maebashi-Kubo} and for transport in a magnetic field in Ref. \cite{Varma_RH2022}, where the experimental results in TBG  \cite{Efetov2022} are compared quantitatively with the theory. The properties which are important to measure to further test the theory for TBG  are thermal conductivity which is predicted to be constant in temperature despite the logarithmic enhancement in the specific heat and the specific heat and the  thermoelectric power which are predicted to have the logarithmic in temperature enhancement near criticality.
 
 
Evidence \cite{Cao:2018ef, EAndrei:2021} has been presented by STM measurements that the super-conductivity in TBG is of the d-wave variety. 
 The results of the theory of quantum-criticality above predicted that the single-particle scattering rate near the Fermi-surface is angle-independent \cite{CMV-MFL}. This has been thoroughly tested in angle-resolved photoemission experiments in cuprates \cite{KaminskiPR2005, Ramshaw2021}. d-wave superconductivity requires that the fermions in the particle-particle channel scatter dominantly through angles near $\pi/2$ \cite{MiyakeSV1986}. This is also clearly in view in the extraction of the symmetry of the pairing interaction in ARPES experiments in cuprates \cite{Bok_ScienceADV}. How angle independent scattering in the particle-hole channel is reconciled with d-wave scattering in the pairing channel follows from the vertex coupling the fermions to the collective critical fluctuations \cite{ASV2010} of the quantum xy model which couple to fermions by the vertex $\gamma_t$.  
 \section{Concluding Remarks}
 
 We have been motivated to show the mapping of a the quantum-critical fluctuations of a likely order in TBG as those of
 a quantum xy model by the resistivity linear in $T$ and in $H$ in several experiments  at various special dopings in TBG \cite{Cao:2018ef, Efetov2022}. Similar behavior is observed in one particular TMD bi-layer \cite{Pasupathy2021}.  In TMD, the critical  fluctuations are magnetic. A theory for them also maps to the xy model and  is given separately. (Liang Fu and C.M. Varma - Unpublished (2023)). No other model has been derived for physically realizable models other than the xy  model which has the $\omega/T$ scaling of fluctuations which are necessary to give the observed properties. The Sachdev-Ye model \cite{SachdevYe1993} or its variation, the SYK model \cite{Kitaev:2018} is derived for what is essentially an infinite range spin-glass coupled to fermions. The applicability of such an interesting model to real physical problems may be questioned. Such a model, which is exactly soluble also gives linear in $T$ resistivity. The essential property of such models is also the $\omega/T$ scaling of the critical fluctuations \cite{CMV-MFL}. 
 
It should be stressed that for the case of the cuprates, long-range order is identified by a variety of experiments and its fluctuations are  mappable to those of the xy model. For the heavy-fermion compounds and the Fe-based superconductors, which have been mentioned, the order parameter is obviously of the antiferromagnetic variety. If, as is the case of CeCu$_6$ the order parameter is Incommensurate Ising \cite{LohneysenRMP2017}, the model for the phase fluctuations is the xy model. For other heavy fermions and the Fe based compounds, the mapping is to the xy model if they are sufficiently anisotropic so that planar order is favored near criticality. Actually, when the critical fluctuations have a dynamical critical exponent $z$ which is effectively infinite as in the quantum xy model, the slightest anisotropy may be enough to have criticality of the xy variety. Direct evidence for order is lacking for the compounds in Moire  bi-layer graphene. Part of the problem is the inapplicability of many of the tools used in bulk material to convincingly decipher the order parameter in materials made up of a few layers.
However, in Ref. \cite{Yazdani2023},  through careful analysis of STM data, evidence for the order we have assumed on the basis of calculations in Ref \cite{Zaletel2020}  is presented for strain-free TBG. For strained samples, a kekule order appears to be found in line with
 some calculations \cite{Parker2021}.  
 
  In the cuprates, the imaginary part of the single particle self-energy is measured to vary as $|\omega|$ and the specific heat consistent with that to be $\propto T \ln T$. This is also consistent with the measured specific heat in heavy-fermions near their quantum-criticality and the thermopower which is measured to be $T \ln T$.  In materials under discussion, the specific heat and thermopower should be deduced in experiments for a detailed test of the theory presented here. The critical fluctuations have been directly measured by polarized scattering in a heavy fermion compound \cite{Schroder1, Schroder2} and in an Fe based compound \cite{Inosov2010}. They have been analyzed \cite{SchroderZhuV2015} and have the product form in frquency and momentum variables as in the solution of the quantum xy model coupled to fermions.
 
 {\it Acknowledgements:} CMV thanks Erez Berg for introducing him to the work of Blutinck et al. and his own on the proposed order in TBG and on extenisve discussions about the generator of rotations of the loop-current pattern proposed for it. Discussions of the experimental data with A. Ghiotto, A. Efetov, P, Jarillo-Herrero, 
 A. Yazdani and M. Zaletel were very helpful.
 
%

\end{document}